\documentclass[twocolumn,floatfix]{revtex4-1}%
\usepackage{graphicx}
\usepackage{amsmath}%
\usepackage{amsfonts}%
\usepackage{amssymb}
\usepackage{bm}
\usepackage{color}

\def\k{{ {\bm k} }}

\def\q{{ {\bm q} }}

\def\0{{ {\bm 0} }}

\allowdisplaybreaks[4]

\begin{document}
\title{
Doping effects on electronic states in electron-doped FeSe: Impact of self-energy and vertex corrections 
}
\author{
Youichi Yamakawa, 
Seiichiro Onari,
and Hiroshi Kontani
}


\date{\today }

\begin{abstract}
 The pairing glue of high-$T_{\rm c}$ superconductivity in heavily electron-doped 
(e-doped) FeSe, in which hole-pockets are absent, has been an important unsolved problem.
 Here, we focus on a heavily e-doped bulk superconductor Li$_{1-x}$Fe$_x$OHFeSe 
($T_{\rm c} \sim 40$~K).
 We construct a multiorbital
model beyond the rigid band approximation and analyze the spin and orbital fluctuations by taking both vertex corrections (VCs) and self-energy into consideration.
 Without e-doping ($x=0$), the ferro-orbital order without magnetism in FeSe is reproduced by the VCs.
The orbital order quickly disappears when the hole-pocket vanishes at $x \sim 0.03$. 
 With increasing $x$ further, the spin fluctuations remain small, 
whereas orbital fluctuations gradually increase with $x$ due to the VCs.
 The negative feedback due to the self-energy is crucial to explain experimental phase diagram. 
 Thanks to both vertex and self-energy corrections, 
the orbital-fluctuation-mediated $s_{++}$-wave state appears for a wide doping range, consistent with experiments.

\end{abstract}

\address{
Department of Physics, 
Nagoya University, 
Nagoya 464-8602, 
Japan 
}

\sloppy

\maketitle
 The high-$T_c$ superconducting (SC) state in heavily electron-doped (e-doped) FeSe systems attracts great attention, but its pairing mechanism is still an open question. 
 One of the characteristics of e-doped FeSe is the lack of magnetic order. 
 Bulk FeSe exhibits spontaneous orbital polarization $n_{xz} \ne n_{yz}$ at $T_S = 90$~K, 
whereas no magnetic order occurs down to the SC transition temperature $T_c = 9$~K 
\cite{Bohmer2017_FeSe_review}. 
The SC state has been studied intensively \cite{Bohmer2017_FeSe_review, Sparau2017_FeSe_orbital-selective, resonance1, impurity4, Baek2020_FeSe_NMR, Yamakawa2017_FeSe_pressure, Kreisel2017_FeSe_gap}.
On the other hand, 
 the orbital order is suppressed by only a few-percent e-doping, and instead, a high-$T_c$ SC phase with $T_c \ge 40$~K appears for a wide doping range in various e-doped FeSe compounds, 
such as an ultra-thin FeSe layer on SrTiO$_3$ ($T_c = 40$--$100$~K) 
\cite{Wang2012_FeSeSTO_1st, Lee2014_FeSeSTO_ARPES, Fan2015_FeSe_STM, Zhang2016_FeSeSTO_ARPES, Shi2017_FeSeSTO_phase}, 
K-dosed FeSe ($T_c \sim 40$~K) 
\cite{Miyata2015_KdoseFeSe, Wen2016_KdosedSTO_phase}, 
and intercalated superconductors ($T_c \sim 40$~K) 
\cite{Lu_LiFeOHFeSe_crystal, Du2016_LiFeOHFeSe_STM, Noji2014_FeSe_intercalate, Yan2016_LiFeOHFeSe_QPI, Niu2015_LiFeOHFeSe_ARPES, Ren2017_LiFeOHFeSe_ARPES, Gu2018_LiFeOHFeSe_QPI}. 
 Angle-resolved photoemission spectroscopy (ARPES) and scanning tunneling microscopy (STM) measurements have revealed that the SC gaps on the electron Fermi surfaces (FSs) are fully gapped 
\cite{Fan2015_FeSe_STM, Zhang2016_FeSeSTO_ARPES, Du2016_LiFeOHFeSe_STM,  Lee2014_FeSeSTO_ARPES, Yan2016_LiFeOHFeSe_QPI, Niu2015_LiFeOHFeSe_ARPES,  Ren2017_LiFeOHFeSe_ARPES,  Gu2018_LiFeOHFeSe_QPI}.

 In usual Fe-based superconductors with electron-FSs (eFSs) and hole-FSs (hFSs), strong spin orbital fluctuations coexist in many compounds. 
 This fact means that two kinds of $s$-wave SC states, the $s_\pm$-wave state with sign reversal and the $s_{++}$-wave state without sign reversal, can be mediated by spin and orbital fluctuations, respectively 
\cite{Kuroki, Mazin, Graser, Kontani, Chubukov, Hirschfeld, Yin}.  
 Up to now, much experimental effort has
been devoted to detecting the presence or absence of sign reversal 
\cite{impurity1, impurity2, impurity4, resonance1, resonance2, resonance5}. 
 The recently reported impurity-induced $s_\pm \rightarrow s_{++}$ crossover in Ba(Fe,Rh)$_2$As$_2$ 
\cite{Schilling2016_s++_imp, Ghigo2018_s++_imp}
has clarified the coexistence of sizable repulsive and attractive pairing glues in Fe-based compounds. 

 In e-doped FeSe compounds, in contrast, the top of the hFSs completely sinks below the Fermi level 
\cite{Lee2014_FeSeSTO_ARPES, Miyata2015_KdoseFeSe, Niu2015_LiFeOHFeSe_ARPES}. 
 In spite of its high $T_c$, 
NMR studies have revealed that the spin fluctuations at $T_c$ in e-doped FeSe are considerably weaker than those in undoped FeSe 
\cite{Hrovat2015_eFeSe_NMR}. 
 It is still a significant mystery that the high-$T_c$ state ($T_{\rm c}>40$~K) is realized in e-doped FeSe in spite of its weak spin fluctuation. 
 Therefore, the pairing glue for the high-$T_{\rm c}$ state in e-doped FeSe is still controversial. 
 Up to now, $d$-wave state 
\cite{Saito2011_KFeSe, Maier2011_KFeSe_dwave, Li2016_eFeSe_dwave, Agterberg2017_full_dwave}  
and incipient $s_\pm$-wave state 
\cite{Saito2011_KFeSe, Chen2015_KFeSe_s+-, Mishra2016_eFeSe_s+-} 
have been proposed based on the spin fluctuation theory, 
while $T_{\rm c}$ will not be high.
 In FeSe/SrTiO$_3$, it is expected that strong interfacial electron-phonon coupling increases $T_c$ up to $\sim 60$~K 
\cite{Lee2015_eFeSe_SOI, Rademaker2016_eFeSe_Phonon, Wang2012_FeSeSTO_1st, Lee2014_FeSeSTO_ARPES}. 
 However, $T_{\rm c}\sim 40$~K is realized in (Li,Fe)OHFeSe even in the absence of strong interfacial electron-phonon interaction \cite{Lu_LiFeOHFeSe_crystal}, 
indicating that the main pairing glue originates from electron correlations.

 The present authors investigated the pairing mechanism in e-doped FeSe by focusing on the vertex corrections (VCs) that induce the 
orbital order and fluctuations 
\cite{Yamakawa2017_eFeSe_SC}. 
 It was found that orbital-fluctuation-meditated $s_{++}$-wave SC can occur even in the absence of hFSs.
 This result is consistent with the recent quasiparticle interference (QPI) measurement reported in Ref. \cite{Fan2015_FeSe_STM}, 
while another QPI study indicates sign reversal between inner- and outer-electron FSs 
\cite{Du2016_LiFeOHFeSe_STM, Gu2018_LiFeOHFeSe_QPI}. 
 However, the reason that a high-$T_{\rm c}$ state is realized in various e-doped FeSe families for a very wide doping range ($x = 0.05$--$0.20$) was not explained 
\cite{Shi2017_FeSeSTO_phase, Wen2016_KdosedSTO_phase}. 
 Therefore, further progress on the theory of pairing mechanisms is still necessary. 

 In this paper, we discuss the mechanism of high-$T_{\rm c}$ superconductivity 
in bulk heavily e-doped compound Li$_{1-x}$Fe$_x$OHFeSe ($T_{\rm c}\sim40$K). 
To understand the $x$-$T$ phase diagram with wide superconducting region,
we analyze the model using the 
self-consistent-vertex correction (SC-VC) theory
\cite{Onari2012_FeSCs_SCVC, Yamakawa2017_eFeSe_SC, Yamakawa2016_FeSe_nem},
by incorporating the $x$-dependent self-energy into the theory
\cite{Onari2014_LFAOH_UVC}. 
 At $x=0$, the ferro-orbital order without magnetism in FeSe is reproduced.
 With increasing $x$, the orbital order quickly disappears, and spin fluctuations remain small for $0.05<x<0.20$.
 Interestingly, orbital fluctuations start to increase gradually for $x>0.06$.
 Therefore, the orbital-fluctuation-mediated $s_{++}$-wave state appears for a wide doping range, in agreement with experiments. 
 The $x$-dependent self-energy is crucial to explain the appropriate e-doping phase diagrams of FeSe compounds.

\begin{figure}[!tb]
\includegraphics[width=1.0\linewidth]{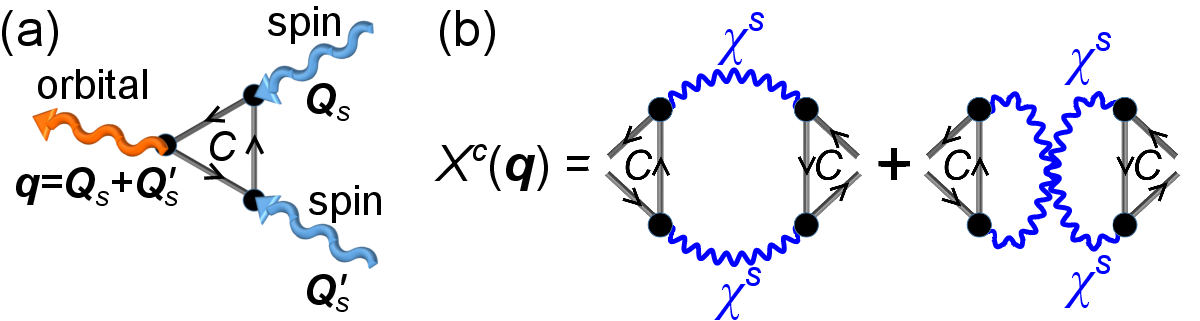}
\caption{
(a) Interference process between spin- and orbital-fluctuations.
$C$ is the three-boson coupling made by electron Green functions.
(b) Charge channel $\chi$-VC $X^{c}$.
}
\label{fig1}
\end{figure}

The key ingredient of the present orbital/spin fluctuation theory
is the interference process between spin- and 
orbital-fluctuations shown in Fig. \ref{fig1}(a) \cite{Tazai2019_CeB6_SCVC}. 
The three-boson coupling $C$ is given by three electron Green functions. 
The correction for the orbital susceptibility
in Fig. \ref{fig1}(b), which we call the $\chi$-VC,
is composed of the process in Fig. \ref{fig1}(a) twice.
Note that the $\chi$-VC at ${\bm q}={\bm 0}$ is proportional to 
$\sum_{\bm{q}}\{\chi^s(\bm{q})\}^2 \propto \max_{\bm{q}}\chi^s(\bm{q})$
in two-dimensional systems \cite{Onari2012_FeSCs_SCVC, Kontani2014_FeSCs_ALVC}.
This ``positive feedback'' through $\chi$-VC
is the physical reason why orbital fluctuations are enlarged 
by spin fluctuations
\cite{Onari2012_FeSCs_SCVC}. 
This bottom-up approach towards strongly correlated systems by introducing the VCs enables us to explain the nematicity and anomalous transport phenomena in Refs. \cite{Onari2012_FeSCs_SCVC, Kontani2008_cuprate_trance}.

Here, we construct the model Hamiltonian for Li$_{1-x}$Fe$_x$OHFeSe. 
We first perform the WIEN2k band calculation for general $x$ 
using the virtual crystal approximation (VCA). 
Next, we derive the eight-orbital tight-binding model 
$\hat{H}_x^{\rm VCA}$ using the Wannier90 package. 
Then, the unfolded eight-orbital $d$-$p$ Hubbard model 
for Li$_{1-x}$Fe$_x$OHFeSe is given as 
$\hat{H} = \hat{H}_x^0 + r \hat{H}^U$, 
where $\hat{H}_x^0 = \hat{H}_x^{\rm VCA}+{\hat \Sigma}^{\rm exp}(\k)$. 
Here, the static self-energy ${\hat \Sigma}^{\rm exp}(\k)$ 
is introduced to reproduce the experimental FSs of FeSe at $x=0$ 
\cite{Malets2014_FeSe_ARPES, Nakayama2014_FeSe_ARPES, Shimojima2014_FeSe_ARPES, Suzuki2015_FeSe_ARPES, Terashima2014_FeSe_deHaas}, 
by following Ref. \cite{Yamakawa2016_FeSe_nem}. 
The microscopic derivation of ${\hat \Sigma}^{\rm exp}(\k)$ 
\cite{GW1, GW2} is an important future issue. 
$\hat{H}^U$ is the first-principles screened Coulomb interaction 
for $d$ orbitals in FeSe (averaged Coulomb interaction is 
$7.2$~eV)  \cite{Miyake2010_FeSCs_DFT}. 
and $r$ is the reduction factor. 
In the present study, we fix $r=0.355$ to 
reproduce weakly developed spin fluctuations for $x=0\sim0.2$ 
above $T_S$ \cite{Wang2016_FeSe_neutron, Pan2017_LiFeOHFeSe_neutron}. 
More detailed explanation is 
summarized in the Supplemental Material (SM) A
\cite{SM}.  

Figures \ref{fig2}(a) and (b) show the folded FSs
in the two-Fe Brillouin zone (BZ) 
derived from $\hat{H}_x$ for $x=0$ and $0.15$, respectively.
Here, we introduced the spin-orbit interaction (SOI) 
$\eta_{\rm SOI} \bm{l} \cdot \bm{s}$ with $\eta_{\rm SOI} = 50$~meV.
At $x = x_c \sim 0.03$, the hFS
around the $\Gamma$ point disappears.
The $d$-orbital density of states (DOS) at the Fermi level
is shown in Fig. \ref{fig2}(c). 
As $x$ increases from $x = 0$, 
both the total DOS and $xz$ orbital DOS decrease for $x \le x_c$.
For $x > x_c$, the $xy$-orbital DOS is dominant and 
it increases gradually with doping. 
The DOS in the FeSe rigid-band approximation (${\hat H}_{x=0}^0$)
is smaller than the DOS in the present VCA-based model,
consistently with a previous study on K-dosed FeSe 
\cite{Choi2019_KFeSe_DFT+VCA+DMFT}. 
This non-rigid-band increment of the DOS 
magnifies $T_{\rm c}$ for $x\sim0.2$.

\begin{figure}[!tb]
\includegraphics[width=1.\linewidth]{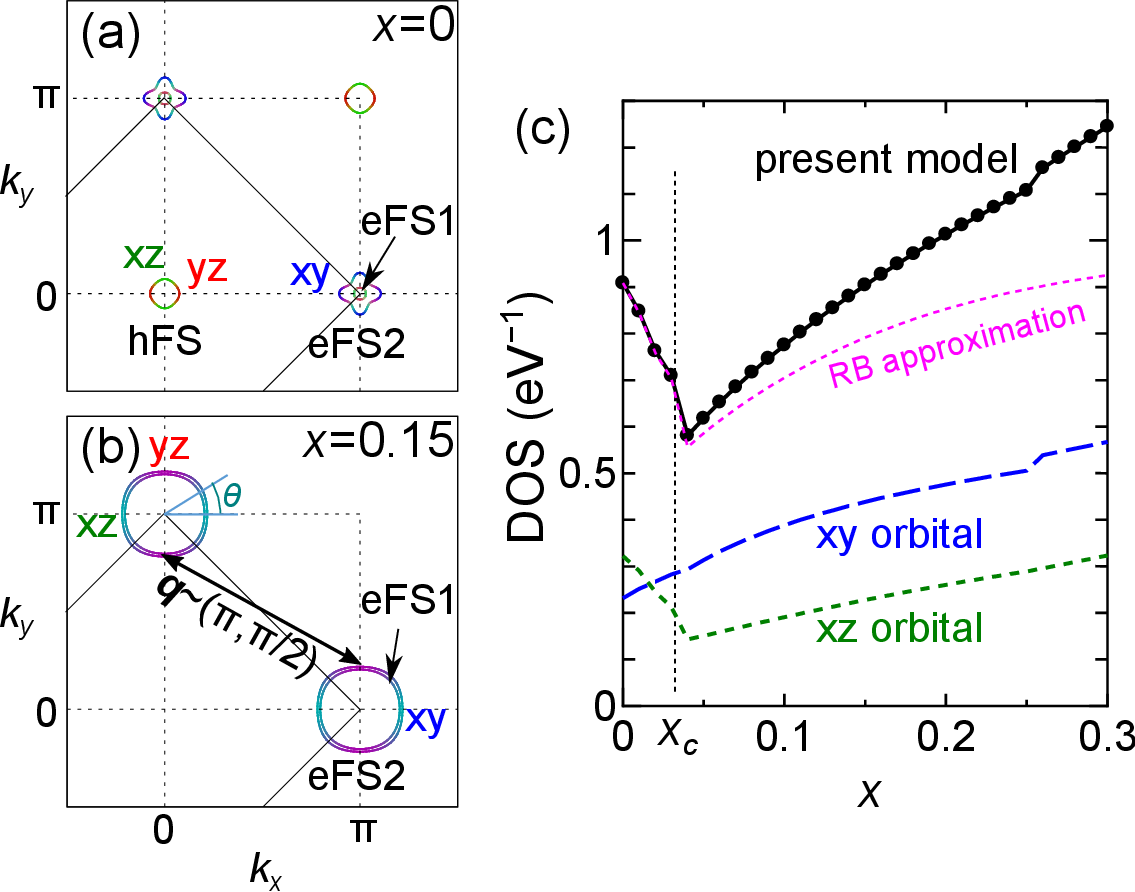}
\caption{
(a)(b) FSs for (a) $x = 0$ and (b) $0.15$
in the folded BZ with $\eta_{\rm SOI} = 50$~meV. 
The green, red, and blue lines correspond to the $xz$, $yz$, and $xy$ orbitals, respectively. 
eFS1 and eFS2 are inner and outer eFSs, respectively. 
$\bm{q} = (\pi, \pi/2)$ is the nesting vector; eFS1 touches eFS2 by the translational wavevector $\bm{q}$.
(c) $x$ dependence of $d$-orbital DOS at the Fermi level. 
The hFS around $\Gamma$ disappears at $x_c \sim 0.03$. 
}
\label{fig2}
\end{figure}

The self-energy 
in the fluctuation exchange (FLEX) approximation \cite{FLEX}
in the absence of the SOI is given as 
\begin{eqnarray}
\Sigma_{l,m} (k) = \frac{T}{N} \sum_{k',l',m'}
V^{\Sigma}_{l,l';m,m'} (k-k')
G_{l',m'} (k'), 
\label{self}
\end{eqnarray}
where $\hat{V}^{\Sigma}=\frac{3}{2} \hat{V}^{s} + \frac{1}{2} \hat{V}^{c}$,
$k = (\bm{k}, \epsilon_n=\pi T(2n+1))$, and
$l,l',m,m'$ represent the $d$-orbital indices. 
($l=$1,2,3,4,5 corresponds to $3z^2-r^2$, $xz$, $yz$, $xy$, $x^2-y^2$ orbitals.)
The electron Green function is 
$\hat{G} (k) = [  \{ \hat{G}^0 (k) \}^{-1} - \hat{\Sigma}(k) ]^{-1}$, 
where $\hat{G}^0(k)$ is the bare Green function.

The spin (charge) susceptibility is
$\hat{\chi}^{s(c)}=\hat{\chi}^{0}[\hat{1}-\hat{U}^{s(c)} \hat{\chi}^{0}]^{-1}$,
where 
$\chi^0_{l,l';m,m'} (q) = - \frac{T}{N} \sum_{k} G_{l,m}(k+q) G_{m'l'}(k)$
is the irreducible susceptibility.
The spin (charge) channel interaction in Eq. (\ref{self}) is 
$\hat{V}^{s(c)} = \hat{U}^{s(c)} + \hat{U}^{s(c)} \hat{\chi}^{s(c)} \hat{U}^{s(c)}$,
where 
$\hat{U}^{s(c)}$ is the spin (charge) channel Coulomb interaction. 
The self-energy represents the mass-enhancement and quasiparticle damping
due to spin fluctuations.
Here, to keep the shape of FSs, 
we drop the static Hermite part in the self-energy
$({\hat\Sigma}(\bm{k},+i\delta)+{\hat\Sigma}(\bm{k},-i\delta))/2$
\cite{Ikeda2010_FLEX, Onari2014_LFAOH_UVC}.

\begin{figure}[!tb]
\includegraphics[width=1.\linewidth]{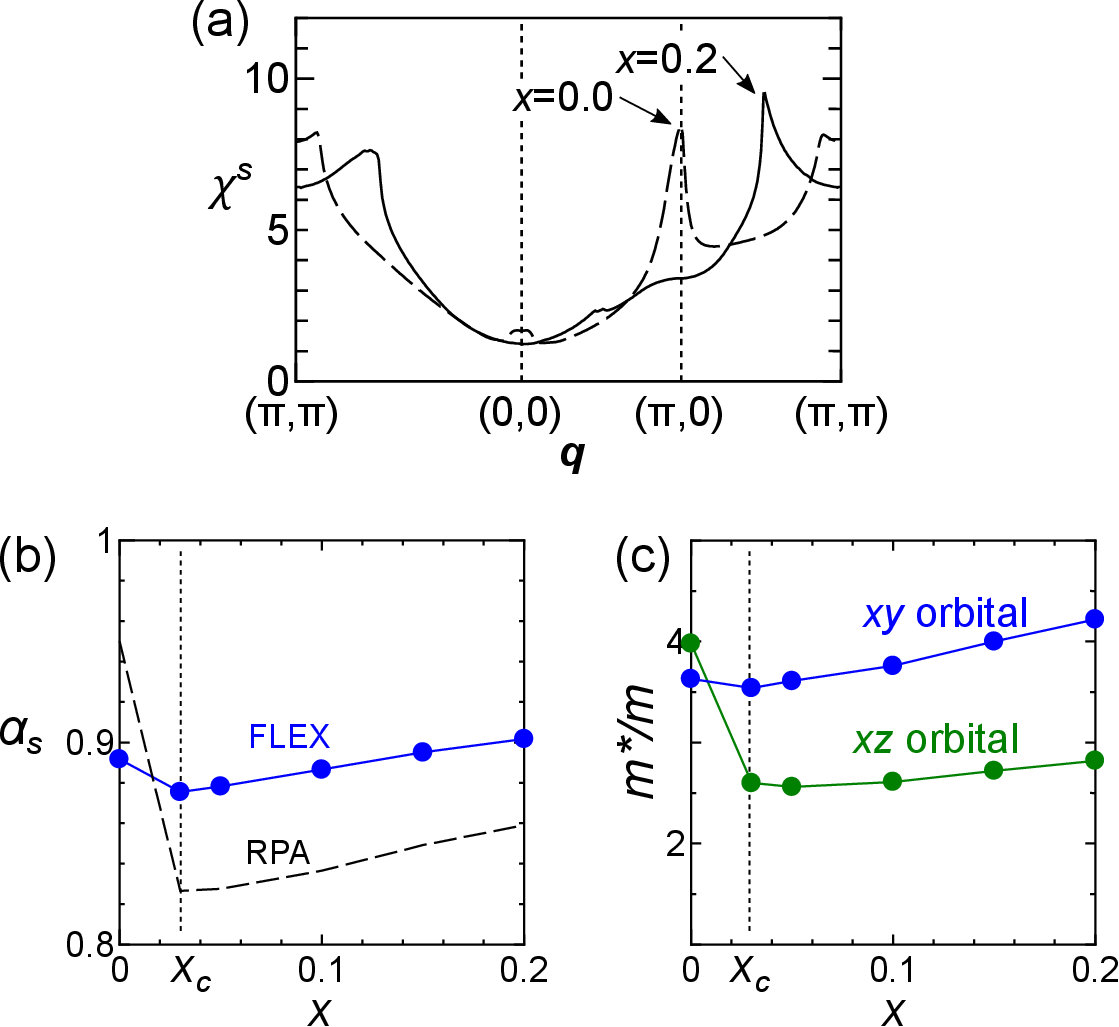}
\caption{
(a--c) Total spin susceptibility $\chi^s (\bm{q})$ for 
(a) $x=0$ and (b) $x=0.2$ for $T = 1$~meV. 
(c) $x$ dependence of spin Stoner factors $\alpha_s$. 
(d) Mass-enhancement factors $Z = m^* / m$ for $xz$ and $xy$ orbitals. 
Here, $r = 0.355$ and $0.218$ are used in the FLEX and RPA calculations, respectively. 
}
\label{fig3}
\end{figure}

Figure \ref{fig3}(a) shows total 
spin susceptibility $\chi^s (\bm{q}) \equiv \sum_{l,m} \chi^s_{l,l;m,m} (\bm{q})$
for $x=0$ and $0.2$ with $r = 0.355$. 
Here, we calculate  $\chi^s$ at low $T$ (=1~meV)
in order to clarify its peak structure.
At $x=0$, $\chi^s (\bm{q})$ has commensurate peaks 
at $\bm{q} = (\pi,0)$ and $(0,\pi)$ due to the nesting
between hFS and eFSs. 
In addition, broad peak appears around $\bm{q} = (\pi,\pi)$. 
At $x=0.2$, $\chi^s (\bm{q})$ has incommensurate peaks
at the nesting vector between eFSs shown in \ref{fig2}(b),
which is observed experimentally
\cite{Pan2017_LiFeOHFeSe_neutron}. 
The importance of the latter nesting has been stressed 
in various FeSe and FeAs compounds in literatures
\cite{Kuroki, Onari2014_LFAOH_UVC, Saito2011_KFeSe, Maier2011_KFeSe_dwave}.

Figure \ref{fig3}(b) shows the $x$ dependence of the spin Stoner factor 
$\alpha_s$, 
which is the maximum eigenvalues of $\hat{U}^s \hat{\chi}^0 (\bm{q})$;
($\alpha_s = 1$ is the magnetic critical point.)
The obtained spin fluctuations remains small 
even for heavily e-doped case, 
which is consistent with NMR results \cite{Hrovat2015_eFeSe_NMR}. 
A similar result is also obtained by the random-phase approximation (RPA)
for $r = 0.209$.
Figure \ref{fig3}(c) shows the mass-enhancement factor 
$Z = m^* / m$ given by the self-energy for $xz \ (yz)$ and $xy$ orbitals. 
$Z_{xy}$ is approximately $3$--$4$ and is larger than $Z_{xz}$ for $x > x_c$, 
due to the strong spin fluctuations and large DOS on the $xy$ orbital.
The obtained values agree with a previous 
dynamical mean-field theory (DMFT) studies 
\cite{Yin2011_mass_DMFT, Choi2019_KFeSe_DFT+VCA+DMFT}.

Next, we study the orbital susceptibility
by including the VC for the susceptibility $\chi$-VC 
($\hat{X}^c$) and $\Sigma$ \cite{Onari2014_LFAOH_UVC}.
Here, we consider the Aslamazov-Larkin (AL) processes
shown in Fig. \ref{fig1}(b) and analytically shown in SM B \cite{SM}, because its significance
in Fe-based superconductors has been discovered in previous studies 
\cite{Onari2012_FeSCs_SCVC, Onari2014_LFAOH_UVC, Yamakawa2016_FeSe_nem, Yamakawa2017_eFeSe_SC, Kontani2014_FeSCs_ALVC, Yamakawa2017_FeSe_pressure}. %
At $x=0$, strong orbital fluctuations with respect to 
$O\equiv n_{xz}-n_{yz}$ appears due to the AL term 
\cite{Onari2012_FeSCs_SCVC,  Yamakawa2016_FeSe_nem}. 
In this case, $\chi^c_{l;m} (\bm{q})\equiv \chi^c_{l,l;m,m} (\bm{q})$
shows large positive (negative) value for $l=m=2$ and $3$
($l=2, m=3$) at $\bm{q}=\bm{0}$,
and they diverge when orbital order $n_{xz}\ne n_{yx}$ appears. 

\begin{figure}[!tb]
\includegraphics[width=1.\linewidth]{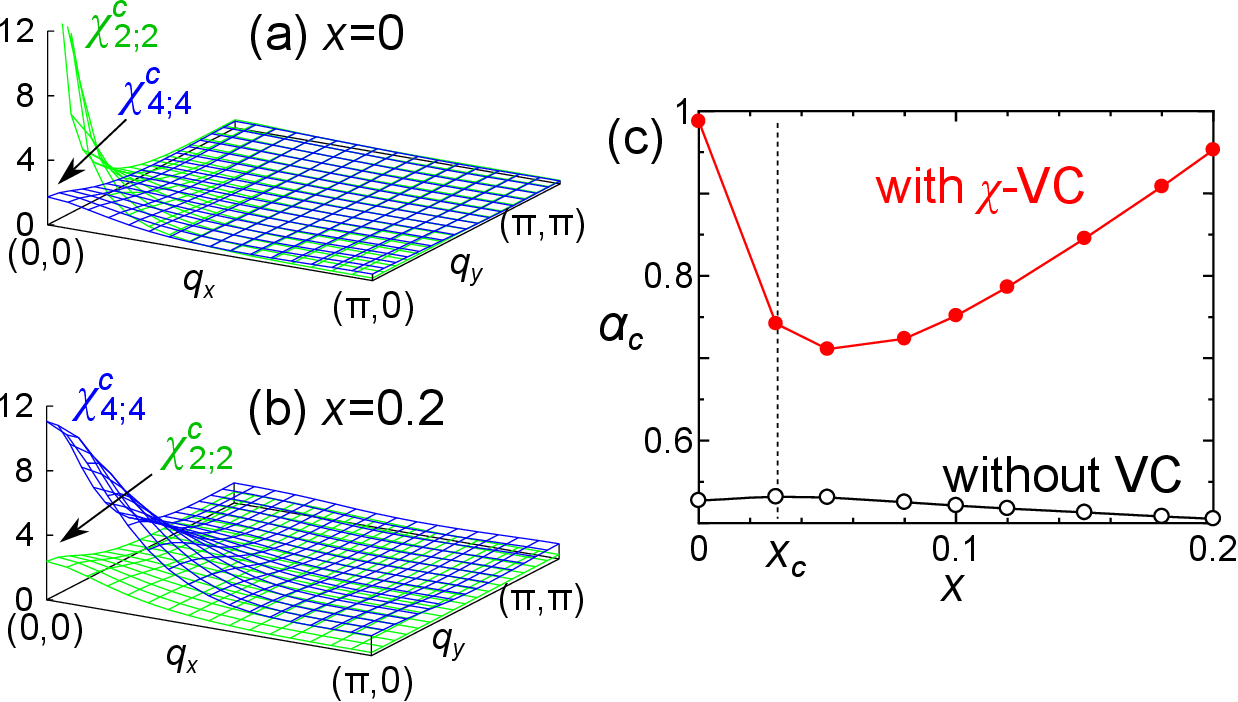}
\caption{
(a) Orbital susceptibilities $\chi^c_{l;l} (\bm{q})$ for 
$l=2=xz$ (green) and $l=4=xy$ (blue) for $x=0$. 
(b) $\chi^c_{l;l} (\bm{q})$ for $x=0.2$.
(c) $x$ dependence of $\alpha_c$ with and without $\chi$-VC. 
Here, we set $r = 0.355$. 
}
\label{fig4}
\end{figure}

Hereafter, we set $T = 20$~meV. 
Figure \ref{fig4}(a) shows the numerical results for $x=0$.
Due to the $\chi$-VC on $xz,yz$ orbitals,
$\chi^c_{2;2}(\bm{0})$
develops divergently due to the $\chi$-VC in spite of 
the weak spin fluctuations in FeSe \cite{Yamakawa2016_FeSe_nem}. 
Since $\chi^c_{2;3} (\bm{0}) \approx -\chi^c_{2;2} (\bm{0})$,
strong ferro-orbital fluctuations with respect to 
$O_{x^2-y^2}\equiv n_{xz}-n_{yz}$ in undoped FeSe
is satisfactorily explained.
The non-magnetic nematic order can be explained in the present theory 
as discussed in Ref. \cite{Yamakawa2016_FeSe_nem} and SM C \cite{SM}. 
In contrast, $\chi^c_{4;4}$ ($4=xy$ orbital) remains small
because the $\chi$-VC on $xy$ orbital is unimportant
due to the smallness of $xy$ orbital spin fluctuations.

In contrast, for $x=0.2$, 
strong ferro-orbital fluctuations 
with respect to $O_{z^2}\equiv n_{xy} - (n_{xz} + n_{yz})/2$ appears.
The obtained $\chi^c_{4;4}(\q)$ is shown in Fig. \ref{fig4}(b),
and the relations
$\chi^c_{2;2} \approx \chi^c_{4;4}/4$,
$\chi^c_{2;4} \approx -\chi^c_{4;4}/2$ and
$\chi^c_{2;3} \approx \chi^c_{4;4}/4$ holds.
As we discussed in Ref. \cite{Onari2014_LFAOH_UVC},
$O_{z^2}$ orbital fluctuations develop when
the $\chi$-VC develop for all $2\sim4$ orbitals.
The crossover of dominant orbital fluctuations
from $O_{x^2-y^2}$ to $O_{z^2}$ occurs at $x\sim x_c$,
reflecting the increment of $xy$-orbital spin fluctuations 
due to inter-eFS nesting.
Both $O_{x^2-y^2}$- and $O_{z^2}$-orbital fluctuations at $\bm{q}\sim \bm{0}$
enlarge $T_c$ irrespective of the gap symmetry 
\cite{Kontani, Kang2016_SC_nematic}.

 Figure \ref{fig4}(c) shows the $x$-dependence of the 
charge Stoner factor $\alpha_c$, which is the maximum eigenvalue of 
$\hat{U}^c ( \hat{\chi}^0 + \hat{X}^{c})$.
Thus, $\alpha_c$ is strongly enlarged by the $\chi$-VC.
Here, $\alpha_c \sim 1$ for $x = 0$ corresponds to the 
ferro-orbital order ($n_{xz} \ne n_{yz}$) in undoped FeSe. 
Through doping, $\alpha_c$ drops quickly since hFS disappears at $x = x_c$. 
Interestingly, $\alpha_c$ gradually increases for $x\gtrsim x_c$,
indicating the orbital-fluctuation-meditated superconductivity
in the absence of the hFS.


To confirm the numerical results in Fig. \ref{fig4},
we also analyze the same model based on the density-wave (DW) equation
developed in Ref. \cite{Onari2016_FeSe_DW, Onari2019_CsFe2As2_B2g, Onari2019_Ba122_AFBO, Kawaguchi-Cu}
in SM D \cite{SM}.
By solving the DW equation, the higher-order AL processes 
are automatically generated.
Also, the criteria of the conserving approximation \cite{Baym}
are satisfied by including the FLEX self-energy.
At $x=0$ and $x=0.25$, strong ferro-orbital fluctuations are obtained,
consistently with Figs. \ref{fig4}(a) and \ref{fig4}(b).
In addition, strong incommensurate charge channel fluctuations
at $\bm{q}\approx(\pi,\pi/2)$ develop at $x=0.25$.
They are overlooked in the SC-VC theory since the 
strong $\bm{k}$-dependence of the form factor,
which represents the fluctuation of hopping integrals, is essential.
This incommensurate ``bond fluctuations''
will be important for the pairing mechanism in heavily e-doped FeSe.


 Next, we study the SC state in e-doped FeSe
by following the theoretical procedure reported in Refs. 
\cite{Yamakawa2017_eFeSe_SC, Saito2015_LiFeAs_SOI} 
 The linearized gap equation is given by 
\begin{eqnarray}
	&& \lambda_{\rm SC} Z_{\alpha} (\bm{k}, \epsilon_{n}) 
	\Delta_{\alpha} (\bm{k}, \epsilon_{n}) \notag \\
	&=& - \frac{\pi T}{( 2 \pi )^2}
		\sum_{m, \beta}
		\oint_{\beta} \frac{d \bm{p}}{v_{\beta}(\bm{p})} 
		V^{\rm SC}_{\alpha, \beta} (\bm{k}, \epsilon_{n}; \bm{p}, \epsilon_{m})
		\frac{\Delta_{\beta} (\bm{p}, \epsilon_{m})}{| \epsilon_{m} |},
\end{eqnarray}
where $\lambda_{\rm SC}$ is the eigenvalue, which is roughly proportional to $T_c$, and the relation $\lambda_{\rm SC} = 1$ is satisfied at $T=T_c$. 
 Further, $\Delta_{\alpha} (\bm{k})$ is the gap function, $v_{\alpha} (\bm{k}) \equiv \frac{\partial \epsilon_{\alpha} (\bm{k})}{\partial \bm{k}}$ is the Fermi velocity, and $Z_{\alpha} (k)$ is the mass-enhancement factor. 
 Here, $\alpha$ and $\beta$ are the indices of the folded FSs with the SOI shown in Figs. 1(a) and (b), 
considering the significance of the SOI on the SC gap. 
 We omit the SOI in calculating the pairing interaction 
since its influence on the fluctuations is small 
\cite{Saito2015_LiFeAs_SOI}.

\begin{figure}[!tb]
\includegraphics[width=.9\linewidth]{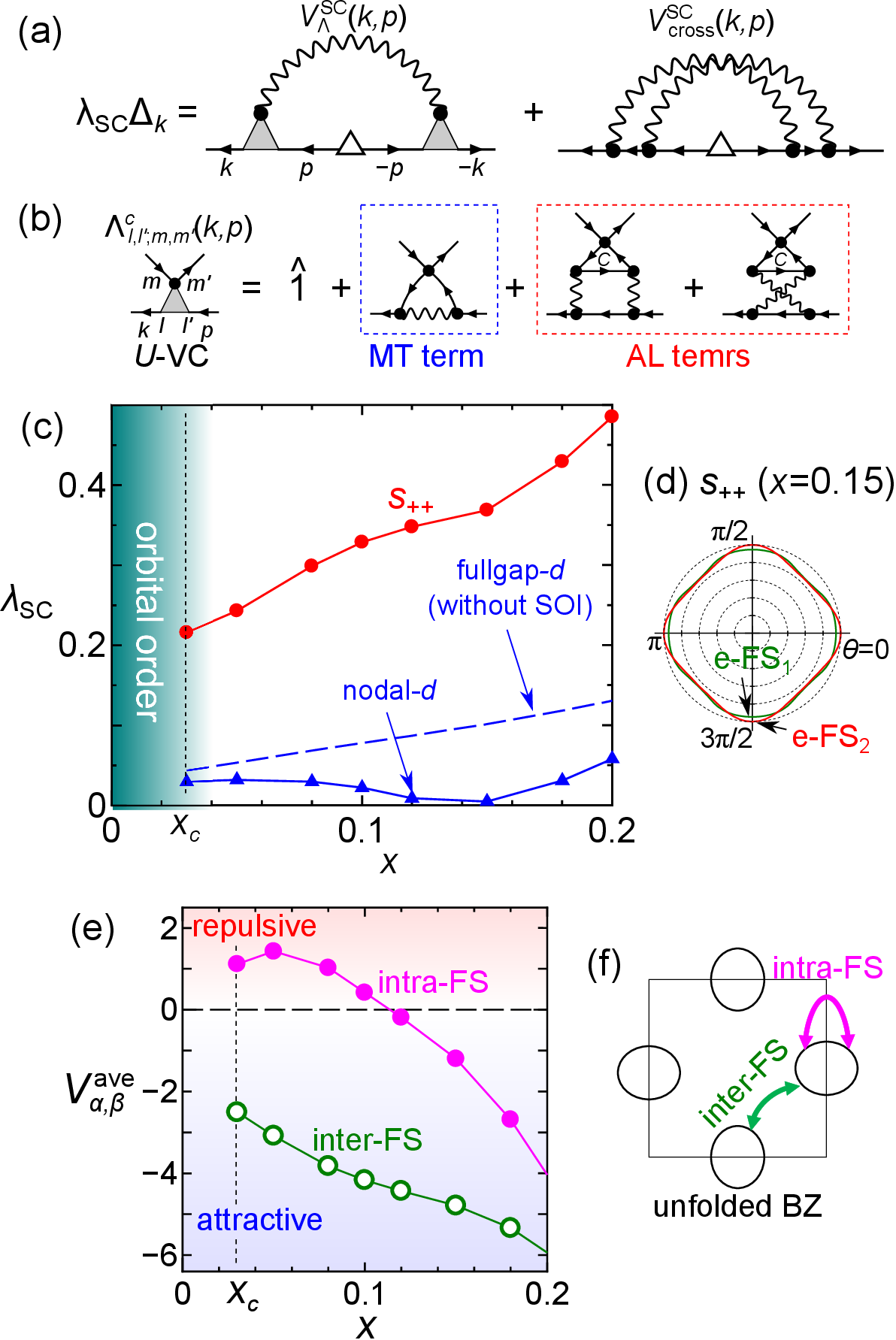}
\caption{
(a) Beyond-ME gap equation in the present analysis.
Both the first term $V^{\rm SC}_{\Lambda}$ and
and the second term $V^{\rm SC}_{\rm cross}$ are significant.
(b) Charge channel $U$-VC $\Lambda^{c}$. 
(c) $x$ dependence of eigenvalues of linearized gap equation $\lambda_{\rm SC}$. 
Here, the SOI $\eta_{\rm SOI} = 50$~meV is considered except for the dashed lines. 
$x < x_c$ is the orbital order region. 
(d) Angle $\theta$ dependence of $s_{++}$-wave gap function on FSs for $x=0.15$. 
(e) Averaged intra-FS ($\alpha = \beta$) and inter-FS ($\alpha \ne \beta$) interactions $V^{\rm ave}_{\alpha, \beta}$ on FSs in (f) unfolded BZ without SOI. 
}
\label{fig5}
\end{figure}

The pairing interaction $V^{\rm SC}$ in the present 
beyond the Migdal-Eliashberg (ME) formalism is shown 
in Fig. \ref{fig5}(a).
The first term is the single-fluctuation exchange process
with the VC for the electron-boson coupling, which we call the $U$-VC.
The $U$-VC $\Lambda^\nu$ ($\nu=s,c$) is 
composed of the AL processes and Maki-Thompson (MT) term 
expressed in Fig. \ref{fig5}(b). 
The first term is symbolically expressed as
$\hat{V}^{\rm SC}_{\Lambda} = \sum_{\nu}^{s,c} b_\nu
\hat{\Lambda}^\mu \hat{V}^\nu \hat{\Lambda}^\nu$,
where $b_s=3/2$ and $b_c=-1/2$.
In the presence of moderate spin fluctuations, $| \Lambda^c |^2  \gg 1$ for low energy region near the Fermi momentum \cite{Tazai2017_CeCu2Si2_SC, Tazai2018_CeCu2Si2_SC, Tazai2016_fRG}. 
Therefore, $U$-VC is significant for the pairing mechanism. 

The second crossing term $\hat{V}^{\rm SC}_{\rm cross}$ 
in Fig. \ref{fig5}(a) is one of the lowest beyond-ME processes
that are absent in the first term.
In Ref. \cite{Yamakawa2017_eFeSe_SC}, we revealed that 
$\hat{V}^{\rm SC}_{\rm cross}$ gives large attractive interaction
between eFS1 and eFS2.
The existence of this inter-pocket attractive pairing 
in heavily e-doped FeSe is confirmed based on the DW equation study 
as we explain in SM E \cite{SM}.
It is found that $\hat{V}^{\rm SC}_{\rm cross}$ represents
the inter-pocket attractive interaction due to the ``bond fluctuations''.

 Figure \ref{fig5}(c) shows the doping dependence of $\lambda_{\rm SC}$. 
 The fully gapped $s_{++}$-wave state has the largest eigenvalue throughout the entire doping region. 
 The resulting $\lambda_{\rm SC}$ for the $s_{++}$-wave state is enhanced
due to the synergy between $U$-VC and $V^{\rm SC}_{\rm cross}$,
as we discuss in Ref. \cite{Yamakawa2017_eFeSe_SC} in detail.
 The resulting fully gapped state with moderate anisotropy 
is shown in Fig. \ref{fig5}(d), which is consistent with experimental reports
in Refs. \cite{Zhang2016_FeSeSTO_ARPES, Lee2014_FeSeSTO_ARPES, Fan2015_FeSe_STM, Niu2015_LiFeOHFeSe_ARPES}. 
The relation with the SC state in bulk FeSe 
 is briefly discussed in SM F \cite{SM}.
In contrast, $\lambda_{\rm SC}$ for the spin-fluctuation-mediated 
$d$-wave state is small, since the spin fluctuations are weak and 
the gap is suppressed by the SOI-induced band mixing 
\cite{Lee2015_eFeSe_SOI}. 
Note that fully-gapped $d$-wave state is realized if 
$\Delta \gtrsim \eta_{\rm SOI}$ \cite{Agterberg2017_full_dwave}. 

 To clarify why the $s_{++}$-wave state is realized,
we show in Fig. \ref{fig5}(e) the $x$ dependence of the averaged 
intra- and inter-pocket interactions:
$V^{\rm ave}_{\alpha, \beta} \equiv \frac{1}{(2 \pi)^2} \sum_{\epsilon_n, \epsilon_m = \pm \pi T} \oint_{\alpha} \frac{d \bm{k}}{v_{\alpha} (\bm{k})} \oint_{\beta} \frac{d \bm{p}}{v_{\beta} (\bm{p})} V^{\rm SC}_{\alpha, \beta} (\bm{k}, \epsilon_n; \bm{p}, \epsilon_m)$.
Here, we consider two electron-pockets in unfolded BZ
without the SOI shown in Fig. \ref{fig5}(f). 
The intra-FS ($\alpha = \beta$) interaction
changes from positive to negative with e-doping
because of the development of the orbital fluctuations shown in Fig. \ref{fig4}(c). 
In addition, sizable attractive inter-FS ($\alpha \ne \beta$) interaction 
is mainly given by $V^{\rm SC}_{\rm cross}$, 
as we discussed in Ref. \cite{Yamakawa2017_eFeSe_SC} and in SM E \cite{SM}.

It is interesting to discuss 
the similarities between e-doped FeSe and heavily e-doped $Re$FeAsO ($Re$ = rare earth),
both of which exhibit high-$T_c$ phases in spite of weak spin fluctuations \cite{Fujiwara2015_LaFeAsOH_NMR, Hrovat2015_eFeSe_NMR}. 
 These high-$T_{\rm c}$ compounds have very similar FSs: large eFSs and tiny or absent hFSs; The FSs in heavily e-doped $Re$FeAsO are shown in Fig. 1(c) in Ref. 
\cite{Onari2014_LFAOH_UVC}.  
 As discussed in Ref. \cite{Onari2014_LFAOH_UVC}, 
the $xy$-orbital DOS is dominant, and weak spin fluctuations on the $xy$-orbital efficiently induce strong orbital fluctuations on the $xy$-orbital, $\chi^c_{xy}$, due to the AL-VCs. 
 The present analysis indicates that similar pairing mechanism is realized in these compounds.

The improvement of the self-energy 
beyond the FLEX approximation is one of the important future issues. 
We note that 
the enhancement of $s$-wave $T_c$ in the Hund's metal state 
was discussed by a recent DMFT study \cite{Fanfarillo}. 

 In summary, we discussed the pairing mechanism in the heavily e-doped bulk compound Li$_{1-x}$Fe$_x$OHFeSe.
 At $x=0$, the ferro-orbital order without magnetism in FeSe is reproduced. 
 With increasing $x$, the orbital order quickly disappears, and the spin fluctuations remain weak for $0<x<0.20$. 
 Interestingly, small spin fluctuations cause large orbital fluctuations 
due to the AL-VC $X^{\rm AL}$ for a wide doping range. 
 Therefore, the orbital-fluctuation-mediated $s_{++}$-wave state appears for $0.05<x<0.2$, consistent with experiments.
 Therefore, the orbital fluctuations will be the main pairing glue in both e-doped FeSe and H-doped 1111 systems.

\acknowledgments
We are grateful to Y. Nomura for fruitful discussion on the doping effect on the band structure in FeSe beyond the rigid-band approximation. 
This work was supported by the Grants-in-Aid for Scientific Research from MEXT, Japan (No. JP19H05825, JP18H01175, JP17K05543, JP17K14338).

\clearpage

\renewcommand{\theHsection}{arabicsection.\thesection}
\renewcommand{\thesection}{\Alph{section}}
\renewcommand{\theHequation}{arabicequation.\theequation}
\renewcommand{\theequation}{S\arabic{equation}}
\renewcommand{\theHfigure}{arabicfigure.\thefigure}
\renewcommand{\thefigure}{S\arabic{figure}}

\makeatother
 \setcounter{figure}{0}
 \setcounter{equation}{0}
 \setcounter{page}{1}
 \setcounter{section}{0}

 \allowdisplaybreaks[4]

\begin{widetext}
\begin{center}
{\bf 
[Supplementary Material] \\
Doping effects on electronic states in electron-doped FeSe: \\
Impact of self-energy and vertex corrections }%
\end{center}
\begin{center}
Youichi Yamakawa, Seiichiro Onari, and Hiroshi Kontani \\
\textit{Department of Physics, Nagoya University, Nagoya 464-8602, Japan}
\end{center}
\end{widetext}


\section{Tight-binding Models}\label{AppendixA}
We explain how we constructed the VCA tight-binding model. 
First, we perform the band calculation of Li$_{1-x}$Fe$_{x}$OHFeSe using WIEN2k. 
Here, we employ the crystal structure of Li$_{0.8}$Fe$_{0.2}$OHFeSe \cite{S-Lu_LiFeOHFeSe_crystal},
and the doping effect is incorporated using the VCA with virtual atoms having a nuclear charge of $3+x$ being substituted for the Li sites. 
Next, we construct the eight-orbital $d$-$p$ tight-binding model by using the Wannier90 and wien2wannier packages for $x = 0$ and $x = 0.2$; other models are obtained via interpolation. 

In many Fe-based superconductors, the results of band calculations differ from the experimental results, especially for FeSe compounds \cite{S-Malets2014_FeSe_ARPES, S-Shimojima, S-Nakayama, S-Suzuki2015_FeSe_ARPES}.
Therefore, to reproduce the experimental Fermi surfaces (FSs), 
we shift the $E_{xz}$ level at $\bm{k}$ = 
[($0, 0$),  ($\pi, 0$),  ($0, \pi$),  ($\pi, \pi$)] by
[-0.24, 0, +0.18, 0]
 and $E_{xy}$ level at $\bm{k}$ =  
[($0,0$),  ($\pi/2,0$),  ($\pi,0$),  ($0,\pi/2$),  ($\pi/2,\pi/2$), ($\pi,\pi/2$), ($0,\pi$), ($\pi/2,\pi$), ($\pi,\pi$)] by
[0, -0.24, +0.4, -0.24, 0, 0, +0.4, 0, -0.3] in unit eV, by introducing the additional inter-orbital hopping integrals for $l$ = $xz$, $yz$, and $xy$ 
\cite{S-Yamakawa2016_FeSe_nem}. 
In addition, we enlarge the hopping parameters between $d_{xy}$ and $p_z$ by 1.25 times to maintain the absence of an eFS at the $\Gamma$ point for $x < 0.15$. 
The present $\bm{k}$- and orbital-dependent shift gives $\hat{\Sigma}_{\rm exp}$ in the main text.

\begin{figure}[!htb]
\includegraphics[width=.99\linewidth]{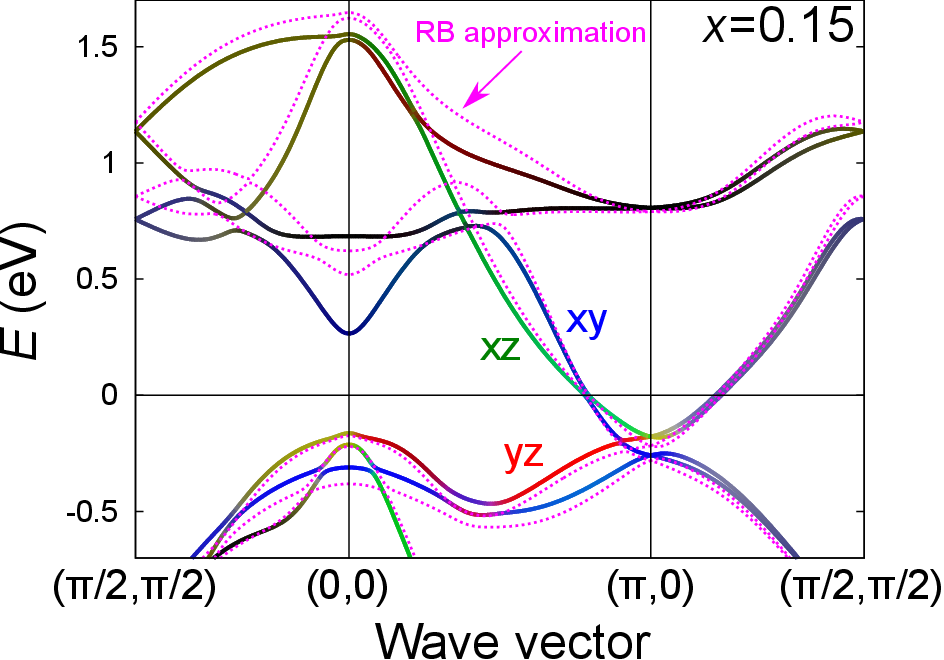}
\caption{
Folded band dispersions for $x = 0.15$ 
with $\eta_{\rm SOI} = 50$~meV. 
The green, red, and blue lines correspond to the $xz$, $yz$, and $xy$ orbitals, respectively. 
The dotted lines represent the band dispersions in the RB model. 
}
\label{fig:S1}
\end{figure}

Figure \ref{fig:S1} shows the folded band dispersions for $x = 0.15$ with the spin-orbit interaction (SOI) $\eta_{\rm SOI} \bm{l} \cdot \bm{s}$ and $\eta_{\rm SOI} = 50$~meV. 
In comparison with the rigid-band (RB) model, the bandwidth obtained using the VCA decreases with doping, similarly to a previous study on K-dosed FeSe \cite{S-Choi2019_KFeSe_DFT+VCA+DMFT}.

The density of states is shown in Fig. \ref{fig2}(c) in the main text, and there is a kink structure at $x \sim 0.25$.  
This kink originates from the emergence of a new electron-pocket around $\Gamma$ point with $e$-doping. 
In Fig. \ref{fig:S1}, we see that the bottom of the electron-like band around $\Gamma$ point is about $+0.25$~eV above the Fermi level 
for $x = 0.15$, whereas it touches to the Fermi level for $x \sim 0.25$. The emergence of this new electron-pocket (= Lifshitz transition) has been reported in heavily e-doped compounds by ARPES studies \cite{S-Shi2017_FeSeSTO_phase, S-Ren2017_LiFeOHFeSe_ARPES}.
Interestingly, $T_c$ increases by $\sim 20K$ at the Lifshitz transition \cite{S-Shi2017_FeSeSTO_phase}. 
It is an important future issue to understand the origin of the increment in $T_c$.

To estimate the value of $\eta_{\rm SOI}$, we make a comparison between the WIEN2k band structure with spin-orbital interaction and the present tight-binding model band structure with $\eta_{\rm SOI} \bm{l} \cdot \bm{s}$. The obtained value is $\eta_{\rm SOI} = 52 \pm 2$~meV, which is very consistent with $\eta_{\rm SOI} = 50$~meV used in the present paper.

\section{Vertex corrections}\label{AppendixB}
Here, we present the analytic expressions for the vertex corrections (VCs) used in the main text. 
In the SC-VC theory, 
the charge susceptibility is given by $\hat{\chi}^{c} (q) = \hat{\Phi}^{c} (q) \left[ \hat{1} - \hat{U}^c \hat{\Phi}^{c} (q) \right]^{-1}$, where $\hat{U}^{c}$ is the charge-channel Coulomb interaction, $\hat{\Phi}^{c} (q) \equiv \hat{\chi}^{0} (q) + \hat{X}^{c} (q)$ is the charge-channel irreducible susceptibility, and $\chi^{0}_{l,l';m,m'} = - \frac{T}{N} \sum_{k} G_{l,m} (k+q) G_{m',l'}(k)$ is the bare susceptibility. 
The charge-channel Aslamazov-Larkin (AL)-VC, $X^{c}$, is given by \cite{S-Onari2012_FeSCs_VC}
\begin{eqnarray}
 	&& X^{c}_{l,l';m,m'} (q) \nonumber \\ 
 	&=& \frac{T}{N} \sum_{p} 
 		\sum_{\scriptsize \begin{array}{c} l_1 \sim l_4, \\ m_1 \sim m_4 \end{array}} 
		{C}^{l_1,l_4}_{l,l_3;l_2,l'} (q,p) {C'}^{m,m'}_{m_1,m_2;m_3,m_4} (q,p) \nonumber \\
 	&& \quad \times \sum_{\nu = s,c} b_{\nu} \hat{V}^{\nu}_{l_1,l_2;m_1,m_2} (p+q) \hat{V}^{\nu}_{l_3,l_4;m_3,m_4} (-p), 
\end{eqnarray}
where $b_s = \frac{3}{2}$ and $b_c = \frac{1}{2}$. 
The three-boson coupling $\hat{C}$ in Fig. \ref{fig1}(a) is given by
\begin{eqnarray}
 	&& {C}^{m,m'}_{m_1,m_2;m_3,m_4} (q,p) = - \frac{T}{N} \sum_{k'} G^0_{m',m_4}(k') \nonumber  \\
 	&& \quad \times G^0_{m_3,m_2}(k'-p) G^0_{m_1,m}(k'+q), 
\end{eqnarray}
and $\hat{C}'$ is defined as 
\begin{eqnarray}
	{C'}^{m,m'}_{m_1,m_2;m_3,m_4} (q,p) \nonumber 
	&& \equiv C^{m,m'}_{m_1,m_2;m_3,m_4} (q,p) \nonumber \\
	&& + C^{m,m'}_{m_1,m_3;m_2,m_4} (q,-p-q). 
\end{eqnarray}

Next, we present the formulations of the $U$-VC. 
The $U$-VC is given as \cite{S-Onari2014_LFAOH_UVC, S-Yamakawa2017_FeSe_SC, S-Tazai2017_CeCu2Si2_SC, S-Tazai2016_fRG, S-Tazai2018_CeCu2Si2_SC}
\begin{eqnarray}
	\hat{\Lambda}^{c(s)} (k,k')= \hat{1}+\hat{\Lambda}^{{\rm MT},c(s) }(k,k') +\hat{\Lambda}^{{\rm AL},c(s)}(k,k'). 
\end{eqnarray}
where $\hat{\Lambda}^{{\rm MT},c(s) }$ is the Maki-Thompson (MT) term, and $\hat{\Lambda}^{{\rm AL},c(s) }$ is the Aslamazov-Larkin (AL) term, both of which were introduced in Refs. \cite{S-Onari2014_LFAOH_UVC, S-Tazai2017_CeCu2Si2_SC, S-Yamakawa2017_FeSe_SC}. 
The charge-channel AL-term is given as
\begin{eqnarray}
	&& \Lambda^{{\rm AL}, c}_{l,l';m,m'} (k,k') \nonumber \\
	&& \quad = \frac{T}{N} \sum_{p} \!\! \sum_{m_1 \sim m_6} \!\! G^0_{m_1,m_2} (k'-p) {C'}^{m,m'}_{m_3,m_4;m_5,m_6} (k-k',p) \nonumber \\
 	&& \quad \times \sum_{\nu = s,c} V^{\nu}_{l,m_1;m_3,m_4} (k-k'+p) V^{\nu}_{m_2,l';m_5,m_6} (-p).
\label{eqn:UALc}
\end{eqnarray}
Here, $\Lambda^{{\rm AL},c}$ and $X^{c}$ have the following relation, 
\begin{eqnarray}
 	&& X^{c}_{l,l';m,m'} (q) \nonumber \\ 
	&=& - \frac{T}{N} \sum_{p} \sum_{m_1,m_2} \Lambda^{{\rm AL}, c}_{l,l';m_1,m_2} (p+q, p) \nonumber \\ 
 	&& \quad \times  G^0_{m, m_1} (p+q) G^0_{m_2, m'} (p).
\end{eqnarray}

In Refs. \cite{S-Tazai2017_CeCu2Si2_SC, S-Tazai2016_fRG, S-Tazai2018_CeCu2Si2_SC}, we verified that $\Lambda^{c}$ is important only for low energy region near the Fermi momentum, and therefore it is important for the pairing interaction. 
We also verified in Ref. \cite{S-Tazai2017_CeCu2Si2_SC} that $\Lambda^{c} \sim O(1)$ if the local approximation is applied. 

Finally, we introduce the crossing-fluctuation-exchange term $\hat{V}^{\rm SC}_{\rm cross}$ \cite{S-Yamakawa2017_FeSe_SC}. 
Its analytic expression is given by
\begin{eqnarray}
	&& \hat{V}^{\rm SC}_{\rm cross} (k, k') =  \frac{T}{4N} \sum_{q} \hat{G}^0 (k'-q) \hat{G}^0 (-k-q) \nonumber \\ 
	&& \quad \times \sum_{\nu, \nu' = c, s} b'_{\nu, \nu'} \hat{V}^{\nu} (k-k'+q) \hat{V}^{\nu'} (-q), 
\end{eqnarray}
where $b'_{c, c}  = -1$ and $b'_{s, s} = b'_{c, s} = b'_{s, c} = 3$.

\section{non-magnetic nematic state in bulk and low-doped FeSe}
 Here, we explain the reason why the orbital order without magnetism can be realized for $x < x_c$ as shown in Fig. \ref{fig4} in the main text.  
 We concentrated on this important issue in Refs. \cite{S-Yamakawa2016_FeSe_nem, S-Onari-form}, and found that the relations $T_S > 0$ (positive nematic transition temperature) and $T_N < 0$ (absence of magnetism) can be satisfied in the present spin-fluctuation-driven orbital order theory. 

Based on the SC-VC theory, when $\chi^s (\bm{Q}) \sim 1/(T-T_N)$, the predicted orbital susceptibility is $\chi^{\rm orb} (\bm{0}) \sim 1/[ 1-C(T) ^2 T \chi^s (\bm{Q}) ]$, where $C(T)$ is the three-boson coupling in Fig. \ref{fig1}(a). 
 Since $C(T)^2 \sim 1/T$ \cite{S-Yamakawa2016_FeSe_nem}, then $C(T)^2 T \equiv g$ is roughly $T$-independent. 
Then, $\chi^{\rm orb} (\bm{0}) \sim (T-T_N)/(T-T_S)$ with $T_S \sim T_N + g$. 
Therefore, $\chi^{\rm orb} (\bm{0})$ diverges at a finite temperature even if $T_N < 0$ once $g > -T_N$. (We stress that, if $C(T)$ is constant, $T_S$ is always negative if $T_N < 0$.) 
 In addition, $T_S$ is enlarged by the finite electron-phonon coupling.

This is supported by the DW equation analysis in Ref. \cite{S-Onari-form}: 
As shown in Fig. 2 (d) of Ref. \cite{S-Onari-form}, spin Stoner factor $\alpha_S$ saturates as a function of $r$ ($\propto U$) for $r > r_c = 0.257$ at a fixed temperature. 
 The origin of the saturation is the poor-nesting caused by the sign-reversing orbital order that is obtained by the DW equation \cite{S-Onari-form}.  
 The sign-reversing orbital order is actually observed by ARPES studies \cite{S-Suzuki2015_FeSe_ARPES}.

 To summarize, because of the (i) $T$-dependence of the three-boson coupling $C(T)$ and (ii) the poor-nesting caused by the sign-reversing orbital order, the relations $T_S > 0$ and $T_N < 0$ can be satisfied in the present theory.

\section{DW equation in conserving approximation}\label{AppendixC}
 Here, we discuss the charge (orbital) order based on the density-wave (DW) equation 
\cite{S-Onari-form, S-Onari2019_Ba122_B2g, S-Onari2019_Ba122_AFBO, S-Kawaguchi-Cu}, 
where the form factor $\hat{f}^{\bm{q}} (\k)$ is taken into account. 
In order to satisfy the conserving approximation (CA) formalism of Baym and Kadanoff \cite{S-Baym}, 
we solve the following DW equation including the FLEX self-energy \cite{S-FLEX}. 
This CA is important to avoid unphysical results.

 The DW equation is given as
\begin{eqnarray}
	&& \lambda_{\bm{q}} f^{\bm{q}}_{l,l'}(k) 
	= \frac{T}{N} \sum_{k'} \sum_{m_1 \sim m_4} I^{c, \bm{q}}_{l,l';m_1,m_2} (k,k') \nonumber \\
	&& \qquad \times G_{m_1, m_3} (k+q) G_{m_4,m_2}(k) f^{\bm{q}}_{m_3,m_4} (k'), 
\label{eqn:linearized}    
\end{eqnarray}
where $\lambda_{\bm{q}}$ is the eigenvalue and $l, m$ are orbital indices.

\begin{figure}[!htb]
\includegraphics[width=.99\linewidth]{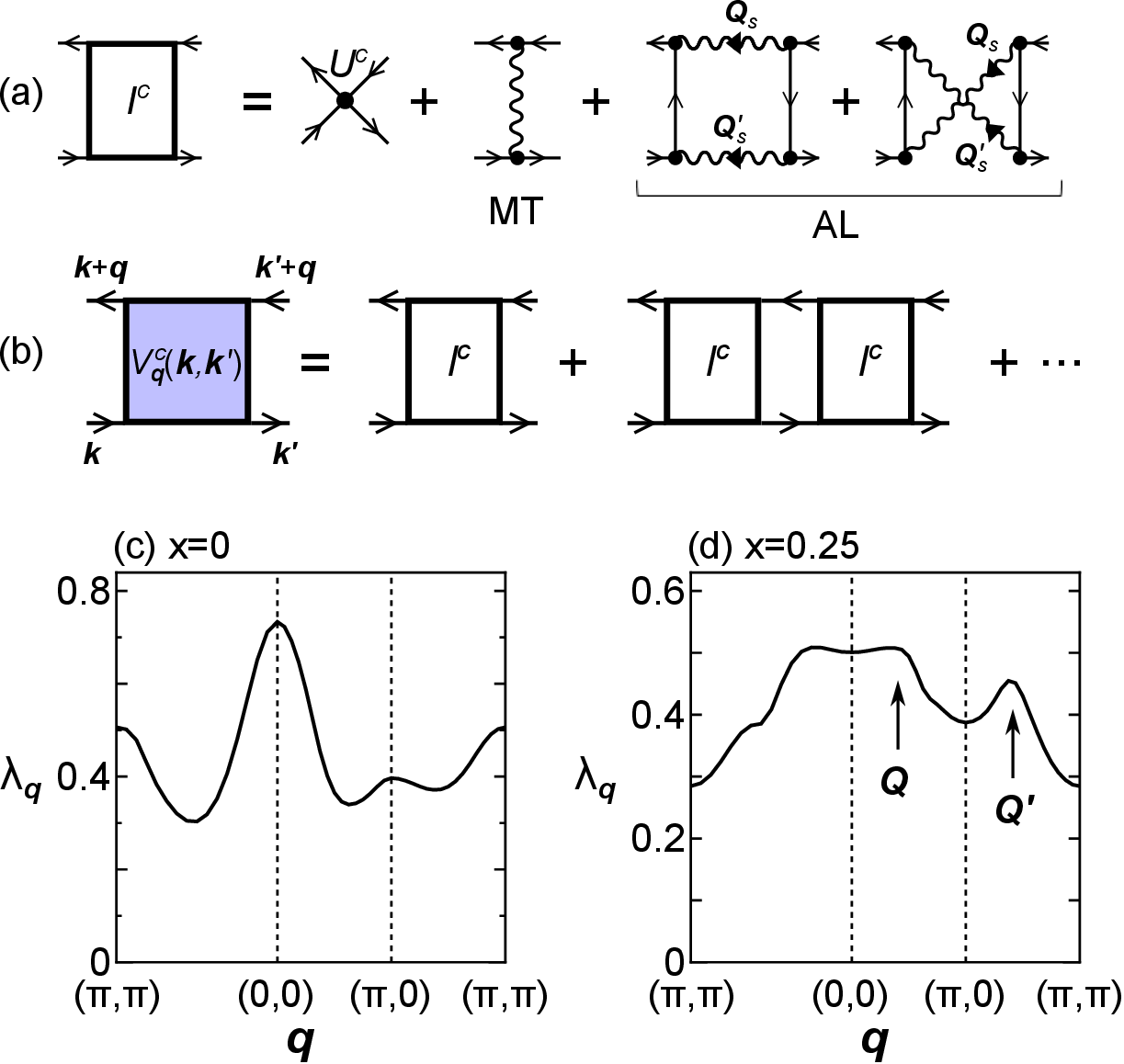}
\caption{
(a) Charge channel irreducible kernel $I^c$ composed of $U^c$, MT term, and AL terms. 
(b) Full four-point vertex.
Momentum dependences of $\lambda_{\bm{q}}$ for (c) $x = 0$ and (d) $x = 0.25$. 
}
\label{dw-fese}
\end{figure}

The diagrammatic expression of the irreducible vertex $\hat{I}$ is shown in Fig. \ref{dw-fese}(a). 
By solving the DW equation, the effective interaction $\hat{V}$ shown in Fig. \ref{dw-fese}(b), which include the infinite series of $\hat{I}$, is generated.

Figure \ref{dw-fese}(c) shows the $\bm{q}$ dependence of the $\lambda_{\bm{q}}$ for $x = 0$ with $r = 0.355$ and $T = 20$~meV. 
There is a sharp peak at $\bm{q} = \bm{0}$ \cite{S-Onari-form}. 
The $\bm{q}$ dependence of $\lambda_{\bm{q}}$ is consistent with the result for $\Sigma = 0$ in Refs. \cite{S-Onari-form, S-Onari2019_Ba122_B2g, S-Onari2019_Ba122_AFBO}.
We find that the condition $\lambda \sim O(1)$ can be realized even when the self energy is included.

$\lambda_{\bm{q}}$ for $x = 0.25$ is shown in Fig. \ref{dw-fese}(d). 
In this case, the peak at $\bm{q} \approx \bm{0}$ becomes broad, and $\lambda_{\bm{q}}$ is large for $| \bm{Q} | \sim \pi/2$, which corresponds to the diameter of each electron FS. 
In addition, $\lambda_{\bm{q}}$ also develops at the inter-FS nesting vector $\bm{Q}' \approx (\pi, \pi/2)$.
Antiferro-bond fluctuations with $\bm{q} \ne \bm{0}$ are expected to be developed in other Fe-based superconductors \cite{S-Onari2019_Ba122_AFBO}.

Here, we make a comparison between the results of DW equation and SC-VC theory. 
The peaks at $\bm{q} \sim \bm{0}$ for $x = 0$ and $x = 0.25$ are consistent with the ferro orbital fluctuations of SC-VC theory. 
On the other hand, the peak at $\bm{q} \sim \bm{Q}'$ for $x = 0.25$, which does not exist in the SC-VC theory, is the bond fluctuation, in which $\bm{k}$ dependence of $f^{\rm DW}_{\bm{q}} (\bm{k})$ is essential. 
We will discuss it in SM. \ref{AppendixD} in detail. 

\section{The Cross Term $V^{\rm SC}_{\rm cross}$: Bond Fluctuation Contribution}\label{AppendixD}
To understand the two pairing interactions in Fig. \ref{fig5}(a) in a unified way,
we construct the charge-channel effective interaction 
$V_q^c(k,k')$ composed of the irreducible kernel function
$I_q^c(k,k')$ shown in Fig. \ref{dw-fese}(a).
Below, we neglect 
the orbital indices to simplify the equation, the Bethe-Salpeter equation for $V_q^c(k,k')$ is
\begin{eqnarray}
	V_q^c(k,k') &=& I_q^c(k,k') \nonumber \\
	&& + \frac{T}{N} \sum_{p} V_q^c(k,p) G(p+q) G(p) I_q^c(p,k'). 
\end{eqnarray}
Its diagrammatic expression is shown in Fig. \ref{dw-fese} (b).
Here, the kernel is well approximated as
$I_q^c(k,k')\approx g f_q(k)f_q(k)^*$,
where $f_q(k)$ is the form factor of the DW equation.
Then, the effective interaction $V_q^c(k,k')$ is expressed as
\begin{eqnarray}
	V_q^c(k,k')\approx g \frac{f_q(k) f_q^*(k')}{1-\lambda_q}, 
\end{eqnarray}
which takes huge value when $1-\lambda_q\lesssim 1$.

\begin{figure}[!htb]
\includegraphics[width=.90\linewidth]{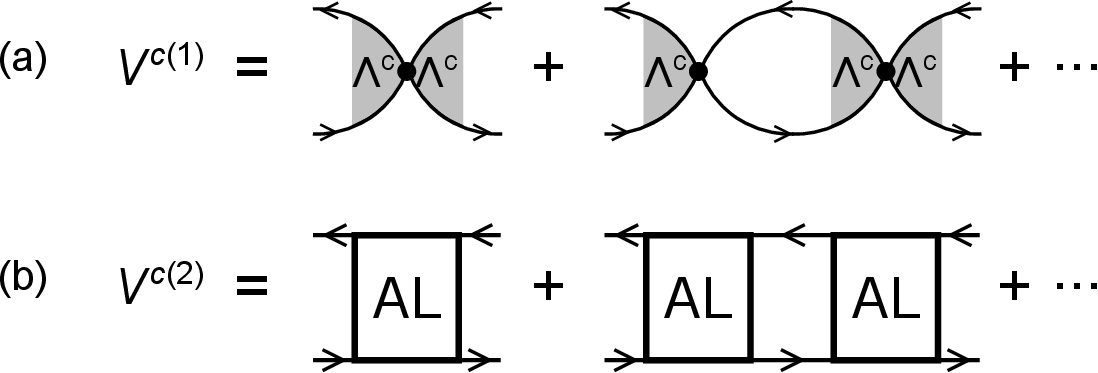}
\caption{
(a) $V^{c(1)}$: the set of the reducible diagrams with respect to $U^c$ in $V$.
(b) $V^{c(2)}$: the set of irreducible diagrams.
}
\label{diagram}
\end{figure}

Here, we divide $V_q^c$ into $V_q^{c(1)}$ and $V_q^{c(2)}$:
The former is the set of the reducible diagrams with respect to $U^c$ in Fig. \ref{diagram}(a), 
and the latter is the set of irreducible diagrams in Fig. \ref{diagram}(b). 
First, we examine the singlet pairing interaction given by $V_q^{c(1)}$; 
$V^{\rm SC(1)}(k,k') \equiv V_{k-k'}^{c(1)}(k,-k')$. 
It corresponds to the first term in Fig. \ref{fig5}(a), $V^{\rm SC}_\Lambda(k,k')$, 
if $\Lambda^c$ in Fig. \ref{diagram}(a) is given as Fig. \ref{fig5}(b). 
Thus, $V^{\rm SC(1)}(k,k')$ represents the orbital-fluctuation-mediated pairing interaction dressed by two $U$-VCs. 

If the form factor has sign reversal and 
$\frac1N \sum_\k f_q(k)\approx 0$ is satisfied, 
the relation $V_q^{c(2)} \gg V_q^{c(1)}$ is realized. 
In this case, the pairing interaction 
$V^{\rm SC(2)}(k,k') \equiv V_{k-k'}^{c(2)}(k,-k')$ becomes significant. 
Its lowest term corresponds to the second term in Fig \ref{fig5} (b), 
$V^{\rm SC}_{\rm cross}(k,k')$, which gives the large attractive inter-pocket interaction for $x=0.20$.
The Fourier transformation of $f_q(k)$ gives the bond-order, 
which is the modulation of hopping integrals. 
Thus, $V^{\rm SC(2)}(k,k')$ represents the bond-fluctuation-mediated pairing interaction.

According to the DW equation analysis for the e-doped FeSe model in Fig. \ref{dw-fese}, 
strong ferro-fluctuations emerge for both $x=0$ and $x=0.25$, 
consistently with the ferro-orbital fluctuations in Figs. \ref{fig4}(a) and \ref{fig4}(b). 
In addition, strong incommensurate fluctuations 
at $\bm{q}\approx(\pi,\pi/2)$ develop at $x=0.25$. 
They are overlooked in the SC-VC theory since the strong $\bm{k}$-dependence of the form factor is essential. 
This incommensurate ``bond fluctuations'' (= fluctuation of hopping integrals)
is important for the pairing mechanism in heavily e-doped FeSe, 
as we see in Fig. \ref{fig5}(d).

\section{Superconductivity in bulk and low-doped FeSe}
Here, we briefly discuss the superconductivity for $x \le x_c$. 
For $x < x_c$, both the spin fluctuation-mediated $s_{\pm}$-wave state \cite{S-Kreisel2017_FeSe_SC} and orbital fluctuation $s_{++}$-wave state \cite{S-Yamakawa2017_FeSe_press} have been proposed. 
In both mechanisms, the orbital selective gap function in FeSe $(\Delta_{yz} \gg \Delta_{xz}, \Delta_{xy})$  \cite{S-Sparau2017_FeSe_orbital-selective}
 is reproduced well
by considering the momentum dependent orbital polarization revealed by Refs. \cite{S-Suzuki2015_FeSe_ARPES, S-Onari-form}.
With increasing $x$, the $s_{++}$-wave gap at $x = 0$ can smoothly change to the plain s-wave gap shown in Fig. \ref{fig5}(d). 
On the other hand, the $s_{\pm}$-wave gap at $x = 0$ may change to the ``incipient $s_{\pm}$-wave state'' by assigning the SC gap on the incipient hole-band below the Fermi level \cite{S-Saito2011_KFeSe, S-Mishra2016_eFeSe_s+-, S-Chen2015_KFeSe_s+-}. 
It is our important future issue to understand the experimental smooth crossover of the SC state around $x \sim x_c$.

\end{document}